\documentclass[a4paper, conference]{IEEEtran}
\IEEEoverridecommandlockouts
\usepackage{amsmath,amssymb,amsfonts}
\usepackage{algorithmic}
\usepackage{graphicx}
\usepackage{textcomp}
\usepackage{xcolor}
\usepackage{booktabs}
\usepackage{threeparttable}
\usepackage{cite}
\def\BibTeX{{\rm B\kern-.05em{\sc i\kern-.025em b}\kern-.08em
    T\kern-.1667em\lower.7ex\hbox{E}\kern-.125emX}}
\begin{document}

\title{Improvement Strategies for Few-Shot Learning in OCT Image Classification of Rare Retinal Diseases\\}

\author{Cheng-Yu Tai$^{1}$, Ching-Wen Chen$^{1}$, Chi-Chin Wu$^{1}$, Bo-Chen Chiu$^{1}$,\\
Cheng-Hung (Dixson) Lin$^{1}$,
Cheng-Kai Lu$^{2}$, Jia-Kang Wang$^{3}$, Tzu-Lun Huang$^{3}$\\[1ex]
\small
$^{1}$Department of Electrical Engineering, Yuan Ze University, Taoyuan City, Taiwan 320, R.O.C.\\
$^{2}$Department of Electrical Engineering, National Taiwan Normal University, Taiwan 106, R.O.C.\\
$^{3}$Department of Ophthalmology, Far Eastern Memorial Hospital, New Taipei City, Taiwan 220, R.O.C.\\
}

\maketitle

\begin{abstract}
This paper focuses on using few-shot learning to improve the accuracy of classifying OCT diagnosis images with major and rare classes. We used the GAN-based augmentation strategy as a baseline and introduced several novel methods to further enhance our model. The proposed strategy contains U-GAT-IT for improving the generative part and uses the data balance technique to narrow down the skew of accuracy between all categories. The best model obtained was built with CBAM attention mechanism and fine-tuned InceptionV3, and achieved an overall accuracy of 97.85\%, representing a significant improvement over the original baseline.
\end{abstract}


\section{Introduction}
GAN-based data augmentation has been proposed to alleviate data scarcity in few-shot learning (FSL) scenarios. Nevertheless, it suffers from notable limitations. As reported by Yoo et al. \cite{b1}, the use of CycleGAN \cite{b2} for synthetic OCT image generation resulted in approximately 20\% of the images being deemed unacceptable by ophthalmologists, ultimately leading to suboptimal classification performance. Moreover, the baseline model exhibited limited robustness, especially in its ability to generalize across rare diseases. This emphasizes the need for more reliable generative methods and enhanced model architectures capable of learning effectively from limited data. To mitigate this, we employ U-GAT-IT \cite{b3}, which incorporates attention mechanisms into the generative process to improve image quality. Furthermore, we apply dataset balancing techniques and integrate both Squeeze-and-Excitation (SE) blocks \cite{b4} and the Convolutional Block Attention Module (CBAM) \cite{b5} into a fine-tuned InceptionV3 \cite{b6} backbone to strengthen feature extraction from critical retinal regions.

\section{Methods}
This study builds upon the framework proposed by Yoo et al. \cite{b1}, which utilized CycleGAN-based augmentation and transfer learning to address the few-shot classification problem in diagnosing rare retinal diseases from OCT images. To overcome the limitations in image synthesis quality and class imbalance inherent in the original framework, we propose an enhanced methodology with three major improvements:
(1) Replacing CycleGAN with U-GAT-IT to generate more acceptable pathological OCT images under few-shot constraints; (2) Implementation of data balancing strategies to mitigate the skewed distribution between common and rare disease classes; and (3) Integrating attention mechanisms, including Squeeze-and-Excitation (SE) \cite{b4} and Convolutional Block Attention Module (CBAM) \cite{b5}, into the InceptionV3 \cite{b6} backbone to improve feature localization and representation.The InceptionV3 \cite{b6} model was initialized with ImageNet pre-trained weights and subsequently fine-tuned on the augmented and rebalanced dataset.

\subsection{U-GAT-IT}
\subsubsection{Basic Augmentation}
We applied linear augmentations as described in Yoo's study \cite{b1} to rare diseases \cite{b7}, including: random translation ± 5\%, rotation ± 30°, zoom 0\,--\,20\%, brightness adjustment ± 10\% and horizontal flip.

\subsubsection{Dataset and Training}
Following the method in Yoo's study \cite{b1} ,we selected 2000 normal images from the OCT2017 dataset\cite{b8} as domain A and expanded each rare disease category to 2,000 images by basic augmentation for domain B. We obtained five GAN models, one for each rare disease and each trained for 200,000 iterations.

\subsection{Data-Balanced}
We sampled 5,000 images from OCT2017\cite{b8} for each major class, then combined 3,000 U-GAT-IT generated images with 2,000 augmented images per rare class. To further improve the model’s generalization ability, we expanded all classes to 10,000 images: major classes were randomly selected; DRUSEN was augmented to 10,000 images; and rare diseases combined 2,000 augmented and 8,000 generated images

\subsection{Attention Module}
\subsubsection{Squeeze-and-Excitation (SE) Block}
To enhance our CNN model, we integrated SE blocks \cite{b4} based on the SE-Inception architecture in SE-Net \cite{b4}. This structure served as the baseline, with fine-tuning applied to the final layers and minor modifications, including adjustments to layer configurations and the reduction ratio.

\subsubsection{Convolutional Block Attention Module (CBAM)}
To enhance the performance of the classifier, the Convolutional Block Attention Module (CBAM) \cite{b5} was integrated into fine-tuned InceptionV3 \cite{b6} architecture. The reduction ratio was varied to ensure that the dual attention strategy of CBAM \cite{b5} effectively captures disease-related features.

\section{Results}
\begin{figure}[htbp]
    \centering
    \includegraphics[width=\linewidth]{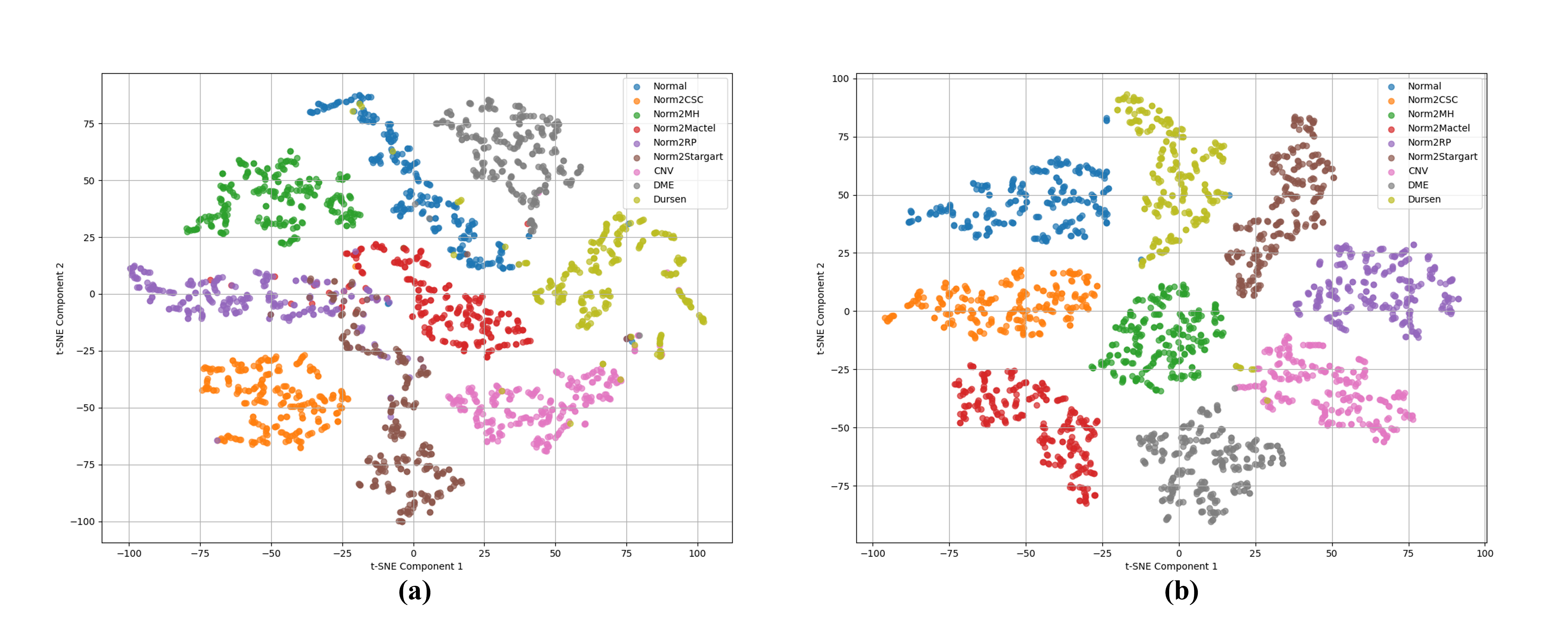}
    \caption{Feature space visualized using the 3D t-SNE technique. (a) t-SNE visualization of GAN-based augmentation by CycleGAN. (b) t-SNE visualization of GAN-based augmentation by U-GAT-IT.}
    \label{fig:t-sne}
\end{figure}

Without any augmentation or balancing, the baseline model (Imb + NoAug) achieved high overall accuracy but low balanced accuracy (0.550), revealing poor performance in detecting rare diseases in Table 1. 
Introducing CycleGAN-based generative augmentation improved balanced accuracy to 0.856 — an increase of approximately 30\% — though this came at the cost of a slight decrease in overall accuracy by about 3.42\%. In contrast, U-GAT-IT outperformed CycleGAN across all metrics; with 5,000 U-GAT-IT generated images (Imb + U5000), the model reached 95.66\% accuracy—roughly 2\% higher than the CycleGAN counterpart (Imb + C5000) in Table 1. The results of the t-SNE \cite{b9} algorithm show that the classification model trained with U-GAT-IT-generated images provides better visualization of rare groups compared to those generated by CycleGAN in Fig. \ref{fig:t-sne}.
After rebalancing all classes to 5,000 images, the CycleGAN-based strategy (Bal + C5000) reduced the accuracy gap between major and average classes from 10.5\% to 3.8\%. In contrast, the U-GAT-IT-based strategy (Bal + U5000) further narrowed the gap from 6.5\% to 4\%, while outperforming the CycleGAN approach by approximately 4\% across all key evaluation metrics.
Under balanced conditions, training with 10,000 U-GAT-IT-generated images (Bal + U10000) the model achieved 96.18\% accuracy, a Cohen’s $\kappa$ of 0.949, and balanced accuracy of 0.959.
Furthermore, incorporating attention mechanisms led to additional improvements. The CBAM\cite{b5}-enhanced model achieved accuracy, Cohen’s $\kappa$, RCI, Matthews correlation, and balanced accuracy of 97.85\%, 0.972, 0.921, 0.972, and 0.972, respectively. Similarly, the SE\cite{b4}-enhanced model (SE) also demonstrated competitive performance, with values of 97.54\%, 0.967, 0.914, 0.967, and 0.926, respectively.

\section{Conclusion}
This paper presents an enhanced framework for few-shot retinal disease classification using OCT images, addressing the limitations of prior generative approaches and feature extraction strategies. Results show that the proposed method achieved an overall accuracy of 97.85\%, marking an approximate 4\% improvement over the baseline model. These findings further confirm that lightweight attention modules, such as CBAM \cite{b5}, can effectively improve classification robustness across both common and rare retinal disease categories.

\begin{table}[htbp]
\centering
\caption{Multiclass performance results pertaining to the nine-class classification of retinal diseases in the five-fold cross-validation}
\resizebox{\linewidth}{!}{%
\begin{tabular}{lccccc}
\hline
\textbf{Method} & \textbf{Accuracy (\%)} & \textbf{Cohen’s $\kappa$} & \textbf{RCI} & \textbf{MCC} & \textbf{BA} \\
\hline
Imb + NoAug & 97.18 ± 0.25 & 0.963 ± 0.003 & 0.964 ± 0.005 & 0.963 ± 0.003 & 0.550 ± 0.001 \\
Imb + C5000 & 93.76 ± 1.44 & 0.922 ± 0.019 & 0.865 ± 0.039 & 0.924 ± 0.019 & 0.856 ± 0.062 \\
Imb + U5000 & 95.66 ± 1.13 & 0.943 ± 0.015 & 0.898 ± 0.019 & 0.944 ± 0.014 & 0.844 ± 0.062 \\
Bal + C5000 & 92.76 ± 1.06 & 0.904 ± 0.014 & 0.778 ± 0.025 & 0.904 ± 0.014 & 0.937 ± 0.010 \\
Bal + U5000 & 95.90 ± 0.38 & 0.946 ± 0.005 & 0.863 ± 0.011 & 0.946 ± 0.005 & 0.977 ± 0.009 \\
Bal + U10000 & 96.18 ± 0.58 & 0.949 ± 0.008 & 0.872 ± 0.016 & 0.949 ± 0.008 & 0.959 ± 0.027 \\
Bal + CBAM & 97.85 ± 0.31 & 0.972 ± 0.004 & 0.921 ± 0.011 & 0.972 ± 0.004 & 0.972 ± 0.020 \\
Bal + SE & 97.54 ± 0.30 & 0.967 ± 0.004 & 0.914 ± 0.008 & 0.967 ± 0.004 & 0.926 ± 0.023 \\
Yoo et al.\cite{b1} & 93.9 ± 4.5 & 0.910 ± 0.065 & 0.969 ± 0.028 & 0.911 ± 0.062  \\
\hline
\end{tabular}
}
\vspace{1mm}
\begin{tablenotes}[flushleft]
\scriptsize
\item 
\textit{Note:} Imb, imbalanced; NoAug, without augmentation; Bal, balanced; Cohen’s $\kappa$, Unweighted Cohen’s kappa; RCI, relative classifier information; MCC, Matthews correlation coefficient; BA, balanced accuracy;  
C5000, CycleGAN (2k real + 3k generated); U5000, U-GAT-IT (2k real + 3k generated); U10000, U-GAT-IT (2k real + 8k generated).
\end{tablenotes}
\end{table}

\begin{table}[htbp]
\centering
\caption{True positive rate per class pertaining to the nine-class classification of retinal diseases in the independent test dataset validation}
\resizebox{\linewidth}{!}{%
\begin{tabular}{lccccccccc}
\hline
\textbf{Method} & \textbf{NORMAL} & \textbf{CNV} & \textbf{DME} & \textbf{Drusen} & \textbf{CSC} & \textbf{MH} & \textbf{MacTel} & \textbf{RP} & \textbf{Stargardt} \\
\hline
Imb + NoAug & 100 & 100 & 99.60 & 96.80 & 0 & 100 & 0 & 0 & 0 \\
Imb + C5000 & 100 & 100 & 98.80 & 99.20 & 80.00 & 100 & 75.00 & 50.00 & 50.00 \\
Imb + U5000 & 100 & 100 & 99.20 & 99.60 & 100 & 100 & 100 & 75.00 & 75.00 \\
Bal + C5000 & 96.00 & 94.20 & 94.20 & 94.90 & 80.00 & 100 & 100 & 100 & 75.00 \\
Bal + U5000 & 98.00 & 96.59 & 96.59 & 97.76 & 100 & 100 & 100 & 100 & 100 \\
Bal + U10000 & 98.60 & 95.90 & 98.98 & 93.63 & 100 & 100 & 100 & 100 & 100 \\
Bal + CBAM & 98.40 & 97.80 & 99.39 & 97.54 & 100 & 100 & 100 & 100 & 100 \\
Bal + SE & 98.40 & 96.99 & 100 & 94.87 & 100 & 100 & 100 & 100 & 75.00 \\
Yoo et al.\cite{b1} & 91.2 & 94.8 & 93.2 & 88.8 & 100 & 100 & 100 & 75 & 75 \\
\hline
\end{tabular}
}
\vspace{1mm}
\begin{tablenotes}[flushleft]
\scriptsize
\item
\textit{Note:} CNV, choroidal neovascularization; DME, diabetic macular edema; CSC, central serous chorioretinopathy;  MacTel, macular telangiectasia;
\end{tablenotes}
\end{table}


\end{document}